\documentstyle[12pt]{article}
\newcommand{\be}{\begin{equation}}
\newcommand{\ee}{\end{equation}}
	\newcommand{\la}{\lambda}
\newcommand{\La}{\Lambda}
\newcommand{\p}{\partial}

\newcommand{\ba}{\begin{array}}
\newcommand{\ea}{\end{array}}
\newcommand{\lb}{\label}
\begin{document}
\setlength{\textwidth}{160mm}
\setlength{\textheight}{230mm}
\renewcommand{\arraystretch}{1.5}
\setlength{\baselineskip}{6mm}
\setlength{\topmargin}{-2.3cm}

\title{The separability and dynamical $r$-matrix for the constrained flows
of Jaulent-Miodek hierarchy}
\author{ Yunbo Zeng\\
Department of Physical Sciences, University of Turku\\
FIN-20500 Turku, Finland\\
Department of Applied Mathematics, Tsinghua University\thanks{Permanent
address.}\\
Beijing 100084, China}

\maketitle
\begin{abstract} We show here the separability of Hamilton-Jacobi
equation for a hierarchy of
integrable Hamiltonian systems
obtained from the constrained flows of the Jaulent-Miodek hierarchy.
The classical Poisson structure
for these Hamiltonian systems is constructed. The associated
$r$-matrices depend not only on the spectral parameters, but also on the
dynamical variables
and, for consistency, have to obey the classical Yang-Baxter equations
of dynamical type. Some new solutions of classical dynamical Yang-Baxter
equations
are presented.
 Thus these integrable systems provide examples both for the dynamical
$r$-matrix and for the separable Hamiltonian system not having a natural
Hamiltonian form.

\end{abstract}

\newpage
\section {Introduction}
The separability of Hamilton-Jacobi equation and classical Poisson
structure for
finite-dimensional intrgrable Hamiltonian system (FDIHS) play an important
role in our
constructing of solutions and understanding of complete integrability for
FDIHS \cite{vi78,l86}.
More resently, interest bas developed in the study of FDIHS admitting
a dynamical $r$-matrix which depends not only on the spactral parameters but
also on the dynamical variables \cite{o90,jm85,ms83}. For example,
the celebrated Calogero-Moser system has been found only recently to
possess a dynamical
$r$-matrix \cite{ja93}. As for the separability, some  FDIHS's were shown
to be
separable in  polar, cartesian, spherical, generalized elliptic and
generalized
parabolic coordinates, respectively,  introduced by using the Lax
representation \cite{eg86,an87,jc94}, or
Painlev\'{e} expansion \cite{vr91}. However, so far all studied separable
Hamiltonian systems have been of the natural form
$H=\frac{1}{2}\sum p_{i}^{2}+V(q_{1},\cdots,q_{n})$.

In this paper we consider a hierarchy of FDIHS's obtained from the
constrained flows \cite{z93a} of the Jaulent-Miodek
hierarchy. The Lax representation for this hierarchy of FDIHS's can be
deduced from the
adjoint representation of the associated eigenvalue problems. By means
of the Lax
representation, we construct the classical Poisson structure for these
FDIHS's. It turns
out that the associated $r$-matrices are of dynamical type, and,
for consistency, have to obey the classical, dynamical Yang-Baxter equations.
Some new solutions of classical dynamical Yang-Baxter equations
are presented.
The Hamiltonian for these FDIHS's is not of the natural form except the
first two members of the hierarchy. We show the
separability of Hamilton-Jacobi equations in new canonical coordinates
introduced via the Lax
representation. Thus these FDIHS's provide examples both for dynamical
$r$-matrix structure and for
separable FDIHS's not having a  natural  Hamiltonian form.

For the following Jaulent-Miodek (JM) eigenvalue problem \cite{mj76}
\be\ba{l}
\left( \begin{array}{c}\psi_{1}\\\psi_{2}\end{array}\right)_{x}=U(u, \lambda)
\left( \begin{array}{c}\psi_{1}\\\psi_{2}\end{array}\right), \\
 U(u, \lambda)
=\left( \begin{array}{cc}0&1\\-\lambda^{2}+\lambda q+r&0\end{array}\right),
\quad \quad
u=\left( \begin{array}{c}q\\r\end{array}\right),\label{a1}\ea\ee
the associated JM hierarchy is of the form
\begin{equation}
\left( \begin{array}{c}q\\r\end{array}\right)_{t_{n}}=J\left
(\begin{array}{c}b_{n+2}\\b_{n+1}\end{array}\right)
=J\frac {\delta H_{n+1}}{\delta u}. \quad
\label{a2}
\end{equation}
Here and after
\begin{equation}\ba{l}
a_{0}=a_{1}=a_{2}=b_{0}=b_{1}=0,\quad b_{2}=-1,\quad b_{3}=-\frac{1}{2}q,
\quad c_{0}=1,\quad c_{1}=-\frac{1}{2}q,\\
\left( \begin{array}{c}b_{m+2}\\ b_{m+1}\end{array}\right)
=L\left( \begin{array}{c}b_{m+1}\\ b_{m}\end{array}\right), \quad
L=\left( \begin{array}{cc}q-\frac{1}{2}\partial_{x}^{-1}q_{x}&
r-\frac{1}{2}\partial_{x}^{-1}r_{x}-\frac{1}{4}\partial_{x}^{2}\\1&0
\end{array}\right),\\
a_{m}=-\frac{1}{2}b_{m,x},\quad c_{m}=a_{m,x}-b_{m+2}+qb_{m+1}+rb_{m},
\quad m=1,2,\cdots,\\
J=\left( \begin{array}{cc}0&2\partial_{x}\\
2\partial_{x}&-q_{x}-2q\partial_{x}\end{array}\right),\quad H_{1}=-q,
\quad H_{m}=\frac{1}{m-1}(2b_{m+2}-qb_{m+1}).
\ea\end{equation}
We have \cite{z93a}
\begin{equation}
\frac {\delta \lambda}{\delta u}=\frac {1}{2}\left( \begin{array}{c}\lambda
\psi_{1}^{2}\\ \psi_{1}^{2}\end{array}\right).
\label{a3}
\end{equation}

The constrained flow of (\ref{a2}) consists of replicas of (\ref{a1}) for $N$
distinct $\lambda_j$ and of restriction of the variational derivatives for
conserved
quantities $H_{k_{0}}$ (for any fixed $k_{0}$) and $\lambda_{j}$
\cite{z91,z94,r92}:
\begin{equation}
\begin{array}{l}
\left( \begin
{array}{c}\psi_{1j}\\\psi_{2j}\end{array}\right)_{x}=U(u, \lambda_{j})
\left( \begin{array}{c}\psi_{1j}\\\psi_{2j}\end{array}\right), \qquad
\quad j=1,...,N,\\
\frac {\delta H_{k_{0}+3}}{\delta u}+
\sum_{j=1}^{N}\frac {\delta \lambda_{j}}{\delta u}=
\left( \begin{array}{c}b_{k_{0}+4}\\b_{k_{0}+3}\end{array}\right)
+\frac {1}{2}\left( \begin{array}{c}<\Lambda \Psi_{1}, \Psi_{1}>\\
<\Psi_{1},\Psi_{1}>\end{array}\right)=0.
\end{array}\label{a4}\end{equation}
Hereafter we denote the inner product in $\bf R^{N}$ by $<.,.>$ and
\begin{equation}
\Psi_1=(\psi_{11},\cdots,\psi_{1N})^{T},\quad\Psi_2
=(\psi_{21},\cdots,\psi_{2N})^{T},\quad \Lambda=
diag (\lambda_1,\cdots,\lambda_N).\end{equation}
It is shown in \cite{z93a} that the system (\ref{a4}) is invariant under
all flows of (\ref{a2}) and
can be transformed into a finite-dimensional integrable
Hamiltonian system (FDIHS) by introducing the so-called
Jacobi-Ostrogradsky coordinates.
The Lax representation for (\ref{a4}), which can be deduced from the
adjoint representation of (\ref{a1}),
is of the form \cite{z93a}
\begin{equation}
M_{x}^{(k_{0})}=[U,M^{(k_{0})}],\label{a6}
\end{equation}
where
\begin{equation}
M^{(k_{0})}\equiv \left( \begin{array}{cc}A^{(k_{0})}&B^{(k_{0})}\\
C^{(k_{0})}&-A^{(k_{0})}
\end{array}\right)=
\left( \begin{array}{cc}A^{(k_{0})}_{1}&B^{(k_{0})}_{1}\\C^{(k_{0})}_{1}&
-A^{(k_{0})}_{1}
\end{array}\right)+N_{0}, \label{a7}\end{equation}
\be
\left( \begin{array}{cc}A^{(k_{0})}_{1}&B^{(k_{0})}_{1}\\C^{(k_{0})}_{1}&
-A^{(k_{0})}_{1}
\end{array}\right)=\sum_{m=0}^{k_{0}+2}\left( \begin{array}{cc}a_{m}&b_{m}\\
c_{m}&-a_{m}\end{array}\right)\la^{k_{0}+2-m},
\end{equation}
\begin{equation}
N_{0}=\frac {1}{2}\sum_{j=1}^{N}\frac{1}{\la-\la_{j}}
\left( \begin{array}{cc}\psi_{1j}\psi_{2j}&-\psi_{1j}^{2}\\
\psi_{2j}^{2}&-\psi_{1j}\psi_{2j}
\end{array}\right).
\end{equation}

We present the first and the third systems of (\ref{a4}) below.

(a) When $k_{0}=0$, we have
\be q=<\Psi_{1},\Psi_{1}>, \quad r=<\La\Psi_{1},\Psi_{1}>-
\frac{3}{4}<\Psi_{1},\Psi_{1}>^{2},\ee
and (\ref{a4}) becomes
\be
\Psi_{1x}=\Psi_{2}=\frac{\p \widetilde{H}_{0}}{\p \Psi_{2}},\quad \quad
\Psi_{2x}=-\frac{\p \widetilde{H}_{0}}{\p \Psi_{1}},\label{a8}\ee
\be\ba{l}
\widetilde{H}_{0}=\frac{1}{2}<\Psi_{2},\Psi_{2}>
+\frac{1}{2}<\La^{2}\Psi_{1},\Psi_{1}>
-\frac{1}{2}<\Psi_{1},\Psi_{1}><\La\Psi_{1},\Psi_{1}>\\
+\frac{1}{8}<\Psi_{1},\Psi_{1}>^{3}.
\ea\ee
The $A^{(0)}_{1},B^{(0)}_{1},C^{(0)}_{1}$ in (\ref{a7}) read
\be \begin{array}{l}
A^{(0)}_{1}(\la)=0,\qquad \quad
B^{(0)}_{1}(\la)=-1,\\
C^{(0)}_{1}(\la)=\la^{2}-\frac{1}{2}<\Psi_{1},\Psi_{1}>\la
-\frac{1}{2}<\La\Psi_{1},\Psi_{1}>+\frac{1}{4}<\Psi_{1},\Psi_{1}>^{2}.
\label{a9}\end{array}\ee
(b) When $k_{0}=2$, by introducing the following Jacobi-Ostrogradsky
coordinates:
\be
q_{1}=q,\qquad q_{2}=r,\qquad p_{1}=-\frac{5}{16}q_{x}q-\frac{1}{8}r_{x},
\qquad
p_{2}=-\frac{1}{8}q_{x},\lb{a10a}\ee
system (\ref{a4}) can be written in canonical Hamiltonian form
\be
\Psi_{1x}=\Psi_{2}=\frac{\p \widetilde{H}_{2}}{\p \Psi_{2}},\qquad
q_{ix}=\frac{\p \widetilde{H}_{2}}{\p p_{i}},\qquad
\Psi_{2x}=-\frac{\p \widetilde{H}_{2}}{\p \Psi_{1}},\qquad
p_{ix}=-\frac{\p \widetilde{H}_{2}}{\p q_{i}},\label{a10} \ee
\be\ba{l}
\widetilde{H}_{2}=-8p_{1}p_{2}+\frac{7}{128}q_{1}^{5}
+\frac{5}{16}q_{1}^{3}q_{2}
+\frac{3}{8}q_{1}q_{2}^{2}+10q_{1}p_{2}^{2}+\frac{1}{2}<\Psi_{2},\Psi_{2}>\\
+\frac{1}{2}<\La^{2}\Psi_{1},\Psi_{1}>
-\frac{1}{2}q_{1}<\La\Psi_{1},\Psi_{1}>-\frac{1}{2}q_{2}<\Psi_{1},\Psi_{1}>.
\ea\ee
The $A^{(2)}_{1},B^{(2)}_{1},C^{(2)}_{1}$ for $M^{(2)}$ are of the form
\be \begin{array}{l}
A^{(2)}_{1}(\la)=-2p_{2}\la-2p_{1}+2p_{2}q_{1},\quad
B^{(2)}_{1}(\la)=-\la^{2}-\frac{1}{2}q_{1}\la-\frac {3}{8}q_{1}^{2}
-\frac{1}{2}q_{2},\\
C^{(2)}_{1}(\la)=\la^{4}-\frac{1}{2}q_{1}\la^{3}-\frac{1}{8}(4q_{2}
+q_{1}^{2})\la^{2}+
\frac {1}{4}(q_{1}^{3}+2q_{1}q_{2}-2<\Psi_{1},\Psi_{1}>)\la\\
-\frac {5}{64}q_{1}^{4}+4p_{2}^{2}+\frac {1}{4}q_{2}^{2}
-\frac {1}{2}<\La\Psi_{1},\Psi_{1}>+\frac {1}{2}q_{1}<\Psi_{1},\Psi_{1}>.
\end{array}\label{a11}\ee
The Hamiltonian $\widetilde{H}_{k_{0}}$ for (\ref{a10}) and for the
FDIHS obtained from (\ref{a4})
with $k_{0}>2$ is not of the natural form. We shall use (\ref{a10}) as
a model to illustrate
how to separate the Hamilton-Jacobi equation for the FDIHS which is not
in the natural Hamiltonian form.
\section {The classical Poisson structure}
In this section we describe the classical Poisson structure associated
with the Lax representation for
(\ref{a8}) and (\ref{a10}).

With respect to the standard Poisson bracket, it is found by a direct
calculation that both
$A^{(0)},B^{(0)},C^{(0)}$ and $A^{(2)},B^{(2)},C^{(2)}$ satisfy the
following relations
\be\begin{array}{l}
\{A(\la),A(\mu)\}=\{B(\la),B(\mu)\}=0,\\
\{C(\la),C(\mu)\}=2g^{(k_{0})}(\la,\mu)(A(\mu)-A(\la)),\\
\{A(\la),B(\mu)\}=\frac{1}{\mu-\la}(B(\mu)-B(\la)),\\
\{A(\la),C(\mu)\}=\frac{1}{\mu-\la}(C(\la)-C(\mu))
+g^{(k_{0})}(\la,\mu)B(\la),\\
\{B(\la),C(\mu)\}=\frac{2}{\mu-\la}(A(\mu)-A(\la)),\end{array}\lb{b1}\ee
with
\be
g^{(0)}(\la,\mu)=<\Psi_{1},\Psi_{1}>-\la-\mu,\quad\qquad
g^{(2)}(\la,\mu)=q_{1}-\la-\mu. \lb{b2}
\ee
We use the standard notation $\sigma_{k}, k=0,1,2,3$ for the Pauli
matrices and $\sigma_{\pm}
=\frac{1}{2}(\sigma_{1}\pm i\sigma_{2})$. In the following let $P$ be
the permutation matrix
\be
P=\frac{1}{2}\sum_{k=0}^{3}\sigma_{k}\otimes\sigma_{k}=
\left( \begin{array}{cccc}1&0&0&0\\0&0&1&0\\0&1&0&0\\0&0&0&1\ea\right).\ee
Denote either $M^{(0)}$ or $M^{(2)}$ by $M$. Set
\be
M_{1}(\la)=M(\la)\otimes I,\qquad\quad M_{2}(\mu)=I\otimes M(\mu).\ee
Then we obtain from (\ref{b1}) that the classical Poisson structure for
both the system
(\ref{a8}) and (\ref{a10}) can be written in the form
\be
\{M_{1}(\la), M_{2}(\mu)\}=[r_{12}(\la,\mu), M_{1}(\la)]
-[r_{21}(\la,\mu), M_{2}(\mu)],\label{b3}
\ee
with the $r$-matrix $r_{12}(\la,\mu)$ given by
\be\ba{l}
r_{12}(\la,\mu)=\frac{1}{\mu-\la}P-g^{(k_{0})}(\la,\mu)S,\qquad\quad
S=\sigma_{-}\otimes\sigma_{-},\\
r_{21}(\la,\mu)=r_{12}^{\pi}(\la,\mu)\equiv Pr_{12}(\mu,\la)P
=r_{12}(\mu,\la). \label{b4}\ea\ee
The $r$-matrix $r_{12}(\la,\mu)$ depends not only on the spectral
parameters but also
on the dynamical variables. The
classical Poisson structure (\ref{b3}) contains all neccesary information
concerning the
considered system and is more rich in content than the Lax representation
\cite{l86}. An
immediate consequence of (\ref{b3}) is that
\be
\{M_{1}^{2}(\la), M_{2}^{2}(\mu)\}=[\overline r_{12}(\la,\mu), M_{1}(\la)]
-[\overline r_{21}(\la,\mu), M_{2}(\mu)],\label{b5}
\ee
where \cite{o90}
\be
\overline r_{ij}(\la,\mu)=\sum_{k=0}^{1}\sum_{l=0}^{1}M_{1}^{1-k}(\la)
M_{2}^{1-l}(\mu)r_{ij}(\la,\mu)M_{1}^{k}(\la)M_{2}^{l}(\mu).\ee
Then it follows from (\ref{b5}) immediately that
\be
4\{TrM^{2}(\la), TrM^{2}(\mu)\}=Tr\{M_{1}^{2}(\la), M_{2}^{2}(\mu)\}=0,
\label{b7}\ee
which esures the involution property of the integrals of motion. For example,
for system (\ref{a10})
one gets
\be
TrM^{2}(\la)=(A^{(2)}(\la))^{2}+B^{(2)}(\la)C^{(2)}(\la)
=-\la^{6}-\widetilde{H}_{2}\la^{2}+F^{(0)}
+\sum_{j=1}^{N}\frac{F^{(j)}}{\la-\la_{j}},\lb{b8}\ee
where
\be\ba{l}
F^{(0)}=-\frac{1}{2}<\La^{3}\Psi_{1},\Psi_{1}>
+\frac{1}{4}q_{1}<\La^{2}\Psi_{1},\Psi_{1}>
+(\frac{1}{4}q_{1}^{2}+\frac{1}{2}q_{2})<\La\Psi_{1},\Psi_{1}>\\
-(\frac{5}{16}q_{1}^{3}+\frac{1}{2}q_{2}q_{1}
-\frac{1}{4}<\Psi_{1},\Psi_{1}>)<\Psi_{1},\Psi_{1}>
-\frac{1}{2}<\La\Psi_{2},\Psi_{2}>\\
-\frac{1}{4}q_{1}<\Psi_{2},\Psi_{2}>
-2p_{2}<\Psi_{1},\Psi_{2}>+4p_{1}^{2}+\frac{5}{2}q_{1}^{2}p_{2}^{2}
-8q_{1}p_{1}p_{2}\\
+\frac{15}{512}q_{1}^{6}
-\frac{3}{32}q_{1}^{2}q_{2}^{2}+\frac{5}{128}q_{1}^{4}q_{2}-2p_{2}^{2}q_{2}
-\frac{1}{8}q_{2}^{3},\\
F^{(j)}=(-2p_{2}\la_{j}-2p_{1}+2q_{1}p_{2})\psi_{1j}\psi_{2j}
+[-\frac{1}{2}\la_{j}^{4}+\frac{1}{4}q_{1}\la_{j}^{3}
+\frac{1}{16}(4q_{2}+q_{1}^{2})\la_{j}^{2}\\
-\frac{1}{8}(q_{1}^{3}+2q_{2}q_{1}-2<\Psi_{1},\Psi_{1}>)\la_{j}
+\frac{5}{128}q_{1}^{4}-2p_{2}^{2}-\frac{1}{8}q_{2}^{2}
+\frac{1}{4}<\La\Psi_{1},\Psi_{1}>\\
-\frac{1}{4}q_{1}<\Psi_{1},\Psi_{1}>]\psi_{1j}^{2}
+(-\frac{1}{2}\la_{j}^{2}-\frac{1}{4}q_{1}\la_{j}-\frac{1}{4}q_{2}
-\frac{3}{16}q_{1}^{2})\psi_{2j}^{2}\\
+\frac {1}{4}\sum_{k\neq j}\frac{1}{\la_{k}-\la_{j}}
(\psi_{1j}\psi_{2k}-\psi_{1k}\psi_{2j})^{2},\qquad\quad j=1,\cdots,N.
\end{array}\ee
Then (\ref{b7}) and (\ref{b8}) guarantee that the functionally independent
integrals of motion $\widetilde{H}_{2}$ and $F^{(j)}, j=0,1,\cdots,N,$
are in involution. This shows the integrability of (\ref{a10}) in the
sense of Liouville \cite{vi78}.

Let us denote
\be\ba{l} \overline {M}_{1}(\la)=M(\la)\otimes I\otimes I,\quad
\overline {M}_{2}(\mu)=I\otimes M(\mu)\otimes I,
\quad\overline {M}_{3}(\nu)=I\otimes I\otimes M(\nu),\\
P_{12}=P\otimes I,\qquad P_{23}=I\otimes P,\qquad P_{13}
=\frac{1}{2}\sum_{k=0}^{3}\sigma_{k}\otimes I\otimes \sigma_{k},\\
S_{12}=S\otimes I,\qquad S_{23}=I\otimes S,\qquad S_{13}
=\sigma_{-}\otimes I\otimes \sigma_{-}.\ea\ee

The Jacobi identity for Poisson bracket can be written as
\be\ba{l}
\{\overline {M}_{1}(\la), \{\overline {M}_{2}(\mu), \overline {M}_{3}(\nu)\}\}
+\{\overline {M}_{3}(\nu), \{\overline {M}_{1}(\la),
\overline {M}_{2}(\mu)\}\}\\
+\{\overline {M}_{2}(\mu), \{\overline {M}_{3}(\nu),
\overline {M}_{1}(\la)\}\}=0.\lb{b9}\ea\ee
One gets the following constraint on the $r$-matrix from (\ref{b9}) \cite{jm85}
\be\ba{l}
[\overline {M}_{1}(\la), [d_{12}(\la,\mu), d_{13}(\la,\nu)]+[d_{12}(\la,\mu),
d_{23}(\mu,\nu)]
+[d_{32}(\nu,\mu), d_{13}(\la,\nu)]\\
+\{\overline {M}_{2}(\mu), d_{13}(\la,\nu)\}
-\{\overline {M}_{3}(\nu), d_{12}(\la,\mu)\}]
+\mbox{cyclic permutations}=0.\lb{b10}\ea\ee
In our case we can show that
\be\ba{l}
d_{ij}(\la,\mu)=\frac{1}{(\mu-\la)}P_{ij}-g^{(k_{0})}(\la,\mu)S_{ij},\\
d_{ji}(\la,\mu)=d_{ij}(\la,\mu),\quad i<j,\quad\qquad i,j=1,2,3, \lb{b12}
\ea\ee
satisfy the following classical Yang-Baxter equations
of dynamical type for the $r$-matrix
\be\ba{l}
 [d_{12}(\la,\mu), d_{13}(\la,\nu)]+[d_{12}(\la,\mu), d_{23}(\mu,\nu)]
+[d_{32}(\nu,\mu), d_{13}(\la,\nu)]\\
+\{\overline {M}_{2}(\mu), d_{13}(\la,\nu)\}
-\{\overline {M}_{3}(\nu), d_{12}(\la,\mu)\}
+\beta^{(k_{0})}[S_{0}, \overline {M}_{2}(\mu)-\overline {M}_{3}(\nu)]
=0,\lb{b11}\ea\ee
plus cyclic permutations, where
\be
\beta^{(0)}=2,\quad\quad \beta^{(2)}=0,\quad\quad S_{0}
=\sigma_{-}\otimes\sigma_{-}\otimes\sigma_{-}.\ee
 So (\ref{b12}) offers solutions to the classical, dynamical
Yang-Baxter equations (\ref{b11}). These equations have an extra term
$[S_{0}, \overline {M}_{i}-\overline {M}_{j}]$ in comparison with the extended
Yang-Baxter equations in \cite{jm85}.

\section {Separability of the Hamilton-Jacobi equation}

By using (\ref{a10}) as a model, we illustrate how the Hamilton-Jacobi
equation can be separated.
In usual way \cite{an87,jc94}, we introduce new coordinates
$u_{1},\cdots,u_{N+2}$ defined by the zeros
of $B^{(2)}(\la)$:
\be
-B^{(2)}(\la)=\la^{2}+\frac{1}{2}q_{1}\la+\frac {3}{8}q_{1}^{2}
+\frac{1}{2}q_{2}
+\frac {1}{2}\sum_{j=1}^{N}\frac{1}{\la-\la_{j}}\psi_{1j}^{2}
=\frac{R(\la)}{K(\la)},\lb{c1}\ee
where
\be
R(\la)\equiv\prod_{k=1}^{N+2}(\la-u_{k})
=\sum_{i=0}^{N+2}(-1)^{i}\beta_{i}\la^{N+2-i},\ee
\be
K(\la)\equiv\prod_{j=1}^{N}(\la-\la_{j})
=\sum_{i=0}^{N}(-1)^{i}\alpha_{i}\la^{N-i},\lb{c2}\ee
and $\alpha_{0}=\beta_{0}=1$,
\be\ba{l}
\alpha_{1}=\sum_{i=1}^{N}\la_{i},\quad \alpha_{2}
=\sum_{i=1}^{N}\sum_{j\neq i}\la_{i}\la_{j},\quad
\beta_{1}=\sum_{i=1}^{N+2}u_{i},\\
\beta_{2}=\sum_{i=1}^{N+2}\sum_{j\neq i}u_{i}u_{j},\cdots.\ea\ee
{}From (\ref{c1}), we obtain
\be
u_{k}^{2}+\frac{1}{2}q_{1}u_{k}+\frac {3}{8}q_{1}^{2}+\frac{1}{2}q_{2}
+\frac {1}{2}\sum_{j=1}^{N}\frac{1}{u_{k}-\la_{j}}\psi_{1j}^{2}=0,
\quad k=1,\cdots,N+2,\lb{c3}\ee
\be
\psi_{1j}^{2}=2\frac{R(\la_{j})}{K'(\la_{j})},\quad j=1,\cdots,N.\lb{c4}\ee
Throughout the paper the prime denotes differentiation with respect to $\la$.

For each $u_{k}$ let us define the additional variable $v_{k}$ as follows
\be
v_{k}=A^{(2)}(u_{k})\\
=-2p_{2}u_{k}+2q_{1}p_{2}-2p_{1}
+\frac {1}{2}\sum_{j=1}^{N}\frac{1}{u_{k}-\la_{j}}\psi_{1j}\psi_{2j},
\quad k=1,\cdots,N+2.\lb{c5}\ee
Notice that $A(\la)=-\frac{1}{2}\partial_{x}B(\la)$, we find from (\ref{c1})
\be
v_{k}=\frac {1}{2}\left(\partial_{x}\frac{R(\la)}{K(\la)}\right)_{\la=u_{k}}=
-\frac{1}{2}\frac{R'(u_{k})}{K(u_{k})}u_{k,x},\quad\quad k=1,\cdots,N+2.
\lb{c6}\ee

We now show that the variables $u_{k}$ and $v_{k}$ are canonical ones:
\be
\{u_{k},u_{j}\}=0,\quad \{u_{k},v_{j}\}=\delta_{kj},\quad \{v_{k},v_{j}\}=0,
\quad k,j=1,\cdots,N+2.\lb{c7}\ee
Equation (\ref{c3}) implies that $u_{k}$ depends only on
$q_{1},q_{2}, \psi_{1j}$, so the first
equality in (\ref{c7}) holds. From (\ref{b1}), one gets
\be\ba{l}
\{B(\la),v_{k}\}=\{B(\la),A(u_{k})\}=\{B(\la),A(\mu)\}_{\mu=u_{k}}
+A'(u_{k})\{B(\la),u_{k}\}\\
=\frac{1}{u_{k}-\la}B(\la)+A'(u_{k})\{B(\la),u_{k}\},\ea\ee
and
\be
0=\{B(u_{j}),v_{k}\}=\{B(\la),v_{k}\}_{\la=u_{j}}+B'(u_{j})\{u_{j},v_{k}\},
\ee
which lead to
\be\ba{l}
\{u_{j},v_{k}\}=\frac{1}{B'(u_{j})}\left(\frac{1}{\la-u_{k}}B(\la)
\right)_{\la=u_{j}}\\
-\frac{A'(u_{k})}{B'(u_{j})}[\{B(u_{j}),u_{k}\}-B'(u_{j})\{u_{j},u_{k}\}]
=\delta_{kj}.\ea\ee
Then it is found from (\ref{b1})
\be\ba{l}\{v_{k},v_{j}\}=
\{A(u_{k}),A(u_{j})\}=\{A(\la),A(\mu)\}_{\la=u_{k},\mu=u_{j}}\\
+A'(u_{j})\{v_{k},u_{j}\}+A'(u_{k})\{u_{k},v_{j}\}
-A'(u_{k})A'(u_{j})\{u_{k},u_{j}\}=0.\ea\ee
This completes the proof of (\ref{c7}).

In order to establish the Hamilton-Jacobi equation in
new coordinates $u_{k}$ and $v_{k}$, we have to express $\widetilde{H}_{2}$
in terms of $u_{k}$ and $v_{k}$.

Multiplying both side of (\ref{c1}) by $K(\la)$ and comparing the
coefficients at $\la^{N+1}, \la^{N},\cdots$, we obtain
\be\ba{l}
q_{1}=2(\alpha_{1}-\beta_{1}),\qquad q_{2}=-2\alpha_{2}
+2\beta_{2}+4\alpha_{1}\beta_{1}-3\beta_{1}^{2}-\alpha_{1}^{2},\\
<\Psi_{1},\Psi_{1}>=2(\alpha_{3}-\beta_{3}-2\alpha_{1}\alpha_{2}
+\beta_{1}\alpha_{2}
+\beta_{2}\alpha_{1}-\beta_{1}\alpha_{1}^{2}+\alpha_{1}^{3}),\lb{c8}
\ea\ee
and similar formular for $<\La\Psi_{1},\Psi_{1}>$ and
$<\La^{2}\Psi_{1},\Psi_{1}>$.
Notice (\ref{a10a}) and $\psi_{2j}=\psi_{1j,x}$, it follows
from (\ref{c4}) and (\ref{c8})
that
\be
\psi_{1j}\psi_{2j}=\frac{R(\la_{j})}{K'(\la_{j})}\sum_{k=1}^{N+2}
\frac{u_{k,x}}{u_{k}-\la_{j}}
=\frac{1}{2}\psi_{1j}^{2}\sum_{k=1}^{N+2}\frac{u_{k,x}}{u_{k}-\la_{j}},
\qquad j=1,\cdots,N,\ee
\be
p_{1}=\frac{3}{8}q_{1}\sum_{k=1}^{N+2}u_{k,x}
+\frac{1}{4}\sum_{k=1}^{N+2}u_{k}u_{k,x},
 \quad\quad \quad
p_{2}=\frac{1}{4}\sum_{k=1}^{N+2}u_{k,x}, \lb{c9}\ee
which together with (\ref{c1}), (\ref{c3}) and (\ref{c6}) gives rise to
the expression of
the terms in $\widetilde{H}_{2}$ containing $p_{1}, p_{2}$ and $\Psi_{2}$:
\be\ba{l}
\frac{1}{2}<\Psi_{2},\Psi_{2}>=\frac{1}{8}\sum_{k=1}^{N+2}
\sum_{l\neq k}\frac{u_{k,x}u_{l,x}}{u_{l}-u_{k}}
\sum_{j=1}^{N}\psi_{1j}^{2}[\frac{1}{u_{k}-\la_{j}}-\frac{1}{u_{l}-\la_{j}}]\\
+\frac{1}{8}\sum_{k=1}^{N+2}u_{k,x}^{2}\sum_{j=1}^{N}\frac{\psi_{1j}^{2}}
{(u_{k}-\la_{j})^{2}}\\
=\frac{1}{4}\sum_{k=1}^{N+2}\sum_{l\neq k}\frac{u_{k,x}u_{l,x}}{u_{l}-u_{k}}
[u_{l}^{2}+\frac{1}{2}u_{l}q_{1}-u_{k}^{2}-\frac{1}{2}u_{k}q_{1}]\\
-\frac{1}{8}\sum_{k=1}^{N+2}u_{k,x}^{2}\left(\frac{d}{d\la}\sum_{j=1}^{N}
\frac{\psi_{1j}^{2}}{\la-\la_{j}}\right)_{\la=u_{k}}\\
=\frac{1}{2}\sum_{k=1}^{N+2}u_{k,x}u_{k}\sum_{l\neq k}u_{l,x}
+\frac{1}{8}q_{1}\sum_{k=1}^{N+2}\sum_{l\neq k}u_{k,x}u_{l,x}\\
+\frac{1}{4}\sum_{k=1}^{N+2}u_{k,x}^{2}[2u_{k}+\frac{1}{2}q_{1}
-\frac{R'(u_{k})}{K(u_{k})}]\\
=8p_{1}p_{2}-10q_{1}p_{2}^{2}-\sum_{k=1}^{N+2}\frac{K(u_{k})}{R'(u_{k})}
v_{k}^{2}.
\lb{c10}\ea\ee
After substituting (\ref{c8}), by a straightforward calculation, the other
terms in $\widetilde{H}_{2}$ becomes
\be\ba{l}
\frac{7}{128}q_{1}^{5}+\frac{5}{16}q_{1}^{3}q_{2}
+\frac{3}{8}q_{1}q_{2}^{2}
+\frac{1}{2}<\La^{2}\Psi_{1},\Psi_{1}>
-\frac{1}{2}q_{1}<\La\Psi_{1},\Psi_{1}>\\
-\frac{1}{2}q_{2}<\Psi_{1},\Psi_{1}>
=-\sum_{i=0}^{5}(-1)^{i}\alpha_{i}\gamma_{5-i},\lb{c11}
\end{array}\ee
where
\be\ba{l}
\gamma_{0}=1,\qquad
\gamma_{1}=\beta_{1},\qquad \gamma_{2}=\beta_{1}^{2}-\beta_{2},\qquad
\gamma_{3}=\beta_{1}^{3}-2\beta_{1}\beta_{2}+\beta_{3},\\
\gamma_{4}=\beta_{1}^{4}-3\beta_{1}^{2}\beta_{2}+\beta_{2}^{2}
+2\beta_{1}\beta_{3}-\beta_{4},\\
\gamma_{5}=\beta_{1}^{5}-4\beta_{1}^{3}\beta_{2}+3\beta_{2}^{2}\beta_{1}
+3\beta_{1}^{2}\beta_{3}-2\beta_{1}\beta_{4}-2\beta_{2}\beta_{3}
+\beta_{5}.\lb{c12}
\ea\ee

In order to express the above formula in terms of $v_{k}$ and $u_{k}$,
we consider an integral
\be
\frac{1}{2\pi i}\oint_{C}\frac{\la^{m}}{R(\la)}d\la=
\sum_{k=1}^{N+2}\frac{u_{k}^{m}}{R'(u_{k})}=R_{\infty}(m),\lb{c13}
\ee
which is taken along the circle centered at origin and with a
sufficiently large radius
to contain all zeros $u_{k}$ of $R(\la)$. The residuum at infinity
$R_{\infty}(m)$ is
equal to a coefficient at $\la^{-1}$ of the expression of
$\frac{\la^{m}}{R(\la)}$.
Then, using (\ref{c12}), (\ref{c13}) lead to
\be
\sum_{k=1}^{N+2}\frac{u_{k}^{m}}{R'(u_{k})}=0,\quad m=0,\cdots,N,\qquad
\sum_{k=1}^{N+2}\frac{u_{k}^{N+m+1}}{R'(u_{k})}=\gamma_{m},\quad m=0,
\cdots,5.\lb{c14}
\ee
Thus , due to (\ref{c10}), (\ref{c11}) and (\ref{c14}), one obtains
\be
\widetilde{H}_{2}=-\sum_{k=1}^{N+2}\frac{1}{R'(u_{k})}
[K(u_{k})v_{k}^{2}+\sum_{i=0}^{5}(-1)^{i}\alpha_{i}u_{k}^{N+6-i}].\lb{c15}
\ee
The Hamilton-Jacobi equation for the action function $S(u_{1},\cdots,u_{N+2})$
 \cite{vi78}
reads (using (\ref{c14}))
\be\ba{l}
\sum_{k=1}^{N+2}\frac{1}{R'(u_{k})}
[K(u_{k})(\frac{\partial S}{\partial u_{k}})^{2}
+\sum_{i=0}^{5}(-1)^{i}\alpha_{i}u_{k}^{N+6-i}]
=h_{1}\\
=\sum_{m=1}^{N+2}h_{m}\sum_{k=1}^{N+2}\frac{u_{k}^{N+2-m}}{R'(u_{k})}.\lb{c16}
\end{array}\ee
It is easy to see that the Hamilton-Jacobi equation (\ref{c16}) may be
separated. Set
$S(u_{1},\cdots,u_{N+2})=\sum_{k=1}^{N+2}S_{k}(u_{k})$, then each
$S_{k}(u_{k})$ has to
satisfy
\be
K(u_{k})\left(\frac{\partial S_{k}}{\partial u_{k}}\right)^{2}
+\sum_{i=0}^{5}(-1)^{i}\alpha_{i}u_{k}^{N+6-i}
-\sum_{m=1}^{N+2}h_{m}u_{k}^{N+2-m}=0,\quad k=1,\cdots,N+2,\lb{c17}
\ee
where $h_{1},\cdots,h_{N+2}$ are the constants of separation. Hence a
complete integral of
Hamilton-Jacobi equation (\ref{c16}) may be written as follows
\be
S(u_{1},\cdots,u_{N+2})=\sum_{k=1}^{N+2}\int^{u_{k}}\sqrt{\frac{G(\la)}
{K(\la)}}d\la,
\lb{c18}\ee
where
\be
G(\la)=-\sum_{i=0}^{5}(-1)^{i}\alpha_{i}\la^{N+6-i}
+\sum_{m=1}^{N+2}h_{m}\la^{N+2-m}.\ee
The constants of separation $h_{1},\cdots,h_{N+2}$ determined by
\be
K(u_{k})v_{k}^{2}+\sum_{i=0}^{5}(-1)^{i}\alpha_{i}u_{k}^{N+6-i}
=\sum_{m=1}^{N+2}h_{m}u_{k}^{N+2-m},\quad k=1,\cdots,N+2,\lb{c19}
\ee
provide a new set of integrals of motion in involution for (\ref{a10}).

In the exactly same way, we can show the separability of Hamilton-Jacobi
equation for
the FDIHS's obtained from (\ref{a4}) in the coordinates $u_{k}$ defined
by the zeros of
$B^{(k_{0})}(\la)$.

\vspace*{7mm}

{\bf Acknowledgments}

The author would like to express his gratitude to Academy of Finland for
financial support and to Professor J. Hietarinta for kind hospitality
and valuable discussions. This work
was also partially supported by Chinese National Basic Research
Project 'Nonlinear Science'.

\end{document}